\documentclass[fleqn,usenatbib]{mnras}

\usepackage{newtxtext,newtxmath}

\usepackage[T1]{fontenc}

\usepackage{amsmath,accents} % amssymb already defined
\usepackage{subcaption,verbatim}%,subfig
\usepackage{graphicx}
\usepackage{color,units}
\usepackage{multirow}
\usepackage{soul}
\usepackage{array, makecell, tabularx, cellspace}

\newcolumntype{L}{>{\raggedright\arraybackslash}X}
\newcolumntype{C}{>{\centering\arraybackslash}X}
\usepackage{booktabs} % for better looking tables
\usepackage{bm}
\usepackage{hyperref}
\usepackage{listings}
\usepackage[normalem]{ulem}
\hypersetup{
    bookmarks=true,         % show bookmarks bar?
    unicode=false,          % non-Latin characters in Acrobat’s bookmarks
    pdftoolbar=true,        % show Acrobat’s toolbar?
    pdfmenubar=true,        % show Acrobat’s menu?
    pdffitwindow=false,     % window fit to page when opened
    pdfstartview={FitH},    % fits the width of the page to the window
    pdfauthor={Boris Goncharov},     % author
    colorlinks=true,       % false: boxed links; true: colored links
    linkcolor=blue,          % color of internal links
    citecolor=blue,        % color of links to bibliography
}
\graphicspath{{figures/}}

\DeclareRobustCommand{\VAN}[3]{#2}
\let\VANthebibliography\thebibliography
\def\thebibliography{\DeclareRobustCommand{\VAN}[3]{##3}\VANthebibliography}

\usepackage{graphicx}	% Including figure files
\usepackage{amsmath}	% Advanced maths commands

\title[Ensemble noise properties of the EPTA]{Ensemble noise properties of the European Pulsar Timing Array} 

\author[Goncharov et al.]{
Boris Goncharov$^{1,2}$\thanks{boris.goncharov@me.com},
Shubhit Sardana$^{1,3}$%,
\\
$^{1}$Max Planck Institute for Gravitational Physics (Albert Einstein Institute), D-30167 Hannover, Germany\\
$^{2}$Leibniz Universität Hannover, D-30167 Hannover, Germany\\
$^{3}$Department of Physics, IISER Bhopal, Bhauri Bypass Road, Bhopal, 462066, India
}

\pubyear{\the\year{}}

\begin{document}
\label{firstpage}
\pagerange{\pageref{firstpage}--\pageref{lastpage}}
\maketitle

% Abstract of the paper
\begin{abstract}
The null hypothesis in Pulsar Timing Array (PTA) analyses includes assumptions about ensemble properties of pulsar time-correlated noise. 
These properties are encoded in prior probabilities for the amplitude and the spectral index of the power-law power spectral density of temporal correlations of the noise. 
Because multiple realizations of time-correlated noise processes are found in pulsars, these ensemble noise properties could and should be modelled in the full-PTA observations by parameterising the respective prior distributions using the so-called hyperparameters. 
This approach is known as the hierarchical Bayesian inference. 
In this work, we introduce a new procedure for numerical marginalisation over hyperparameters. 
The procedure may be used in searches for nanohertz gravitational waves and other PTA analyses to resolve prior misspecification at negligible computational cost. 
Furthermore, we infer the distribution of amplitudes and spectral indices of the power spectral density of spin noise and dispersion measure variation noise based on the observation of 25 millisecond pulsars by the European Pulsar Timing Array (EPTA). 
Our results may be used for the simulation of realistic noise in PTAs. 
\end{abstract}

% Select between one and six entries from the list of approved keywords.
% Don't make up new ones.

%%%%%%%%%%%%%%%%%%%%%%%%%%%%%%%%%%%%%%%%%%%%%%%%%%

%%%%%%%%%%%%%%%%% BODY OF PAPER %%%%%%%%%%%%%%%%%%

\section{\label{sec:intro} Introduction}
Pulsar Timing Arrays~\citep[PTAs,][]{FosterBacker1990} are experiments that monitor pulse arrival times from galactic millisecond pulsars with a primary goal of detecting nanohertz-frequency gravitational waves~\citep{Sazhin1978,Detweiler1979}. 
The most promising source of such gravitational waves is the stochastic superposition of inspiralling supermassive binary black holes in the nearby universe~\citep{RosadoSesana2015}.
Thanks to spacetime metric perturbations from gravitational waves, pulse arrival times experience delays and advances (henceforth, delays). 
The power spectral density (PSD) of delays $P(f)$ induced by stochastic gravitational waves manifests temporal correlations and corresponds to the background's characteristic strain spectrum $h_\text{c}(f)$. 
The Fourier frequency of timing delays $f$ is exactly the gravitational wave frequency.
For the isotropic stochastic gravitational wave background from circular binaries where the inspiral is driven by gravitational wave emission alone, $h_\text{c} \propto f^{-2/3}$ corresponding to $P(f) \propto f^{-13/3}$, highlighting signal prominence towards lower frequencies or, equivalently, longer observational time scales \citep{RenziniGoncharov2022,Phinney2001}.
At low frequencies, PTAs are limited by time-correlated ``red'' noise which is also modelled using a power law. 
The two basic sources of red noise in PTA data are the dispersion measure (DM) variation noise \citep{KeithColes2013} and the spin noise \citep{ShannonCordes2010}.
DM noise features an additional dependence $P(f) \propto \nu^{-2}$, where $\nu$ is a radio frequency.
It belongs to a broader class of noise processes called ``chromatic'', referring to the dependence of the signal amplitude on radio frequency. 
Achromatic red noise that is independent of $\nu$ is also called spin noise because it is associated with irregularities in pulsar rotation. 
A decisive contribution of the stochastic background to PTA data is ultimately determined through the covariance of the signal across pulsar pairs.
The covariance follows the~\citet{HellingsDowns1983} function of pulsar angular separation. 

Contemporary PTA analyses are performed using Bayesian inference, where a uniform prior probability is assumed for pulsar-specific red noise parameters. 
Namely, the log-10 amplitude, $\lg A$, and the spectral index, $\gamma$, of noise PSD. 
Often the term ``prior'' is used to refer to our knowledge of a parameter before evidence is taken into account. 
However, more precisely, priors are the models of how likely a parameter with a certain value is to be found in a given data realization. 
A prior choice is sufficient as long as it is consistent with an observation. 
Uniform priors on PTA noise parameters are designed to be sufficient for single-pulsar noise analyses. 
However, all pulsars represent different realizations of data with respect to pulsar noise parameters\footnote{While remaining a single realization of data with respect to parameters governing $h_\text{c}(f)$ of gravitational wave background.}.  
So, for full-PTA analyses with multiple pulsars, a mismatch between our prior on $(\lg A,\gamma)$ and the observed distribution of these parameters may accumulate across pulsars and lead to biased inference. 
This case is known as \textit{prior misspecification}. 

Implicitly, noise prior misspecification is shown by~\citet{HazbounSimon2020} to bias measurements of the strain amplitude of the gravitational wave background in PTAs. 
The authors have mitigated the bias by allowing the data to choose whether red noise is present or absent in a pulsar. 
Furthermore, in simulations of~\citet{GoncharovShannon2021,ZicHobbs2022}, pulsar spin noise with a wide range of $(\lg A,\gamma)$ modelled with the standard uniform priors, not representative of ensemble noise properties, yields evidence for a stochastic signal with the same power-law PSD across pulsars (which is not present in the simulated data). 
Such a common-spectrum process (CP), under the assumption of uniform noise priors, has been reported in real PTA data as a possible precursor to Hellings-Downs correlations of the gravitational wave background~\citep{GoncharovShannon2021,NG12_GWB,ChenCaballero2021}.

\citet{GoncharovThrane2022} regularised incorrect pulsar noise priors by allowing noise amplitudes of the CP of the putative gravitational wave background to vary across pulsars.
By showing that this variance is consistent with zero, the authors confirmed the consistency of the signal with the gravitational wave background.  
Although the model of \citet{GoncharovThrane2022} provides the working solution and has clear use cases, the ultimate solution to the misspecification of noise priors is to parameterise priors for all relevant noise parameters.
Both the statement of the problem and the most general solution are clearly outlined in~\citet{vanhaasteren2024}. 
The distribution of pulsar noise parameters is inferred simultaneously with a search for Hellings-Downs correlations and other red processes with pulsar-to-pulsar covariance. 
The technique is shown to mitigate systematic errors introduced by incorrect noise priors. 

In this study, we introduce two new methods for modelling ensemble pulsar noise properties in PTA data. 
However, unlike~\citet{vanhaasteren2024}, (1) one of our methods can only be used to infer ensemble pulsar noise properties and (2) another method can only be used to remove a systematic error from incorrect pulsar noise priors without gaining access to ensemble noise properties. 
Next, we perform inference of ensemble noise properties of pulsars from the Second Data Release~\citep[DR2,][]{EPTA_DR2_TIMING} of the European Pulsar Timing Array~\citep[EPTA,][]{DesvignesCaballero2016, KramerChampion2013}.

The rest of the paper is organized as follows. 
In Section~\ref{sec:methodology}, we outline the data analysis methodology and an overview of sources of noise in PTAs. 
Our two new methods of modelling ensemble pulsar noise properties are presented in Sections~\ref{sec:hierarchical:reweigh} and~\ref{sec:hierarchical:marg}.
In Section~\ref{sec:results}, we report on our main results, where each subsection corresponds to a different model of a distribution of noise parameters $(\lg A,\gamma)$. 
In Section~\ref{sec:doubledipping}, we test the robustness of our models to circular analysis. 
Finally, in Section~\ref{sec:conclusion}, we draw conclusions. 

\section{Methodology}\label{sec:methodology}

In this study, we analyze the second data release (DR2,~\citet{EPTA_DR2_TIMING}) of the EPTA. 
In particular, we focus our attention on the ``DR2full'', which we will thus refer to as EPTA DR2 or data unless specified otherwise. 
The data is based on pulse arrival times from a set of 25 millisecond radio pulsars observed over time spans varying between 14 and 25 years. 
The data also includes pulsar timing models obtained with the least-squared fitting of pulse arrival times to the models~\citep{HobbsEdwards2006}. 
Timing models describe how pulse arrival times are affected by deterministic properties of individual pulsars such as pulsar spin frequency and derivatives, pulsar position and proper motion in the sky, dispersion measure (DM) and derivatives, and binary orbital parameters (if applicable). 
Contributions of other signals and noise processes to data are referred to as ``residuals'', implying that they yield a difference between pulse arrival times predicted by the timing model and measured arrival times. 
Pulse arrival times are referenced to the position of the Solar System barycenter which is defined based on the DE440 ephemeris~\citep{ParkFolkner2021}.

\subsection{\label{sec:standard_methodology} Standard PTA data analysis methodology}

Contributions to the PTA data beyond the timing model are often determined using Bayesian inference as described below. 
The likelihood of data $\bm{\delta t}$ (a vector of the measured pulse times of arrival, ToA) is a Gaussian distribution which is multivariate with respect to a number of observations, 
\begin{equation}
    \mathcal{L}(\bm{\delta t} | \bm{\theta}) = \frac{\exp\bigg(  -\frac{1}{2} (\bm{\delta t} - \bm{\mu})^T \bm{C}^{-1}(\bm{\delta t}-\bm{\mu}) \bigg)}{\sqrt{\det\left( 2\pi \bm{C}\right)}},
\end{equation}
where $\bm{\theta}$ is a vector of parameters of models that describe the data.
In other words, $\mathcal{L}(\bm{\delta t}| \bm{\theta})$ is the time-domain likelihood. 
The model prediction for pulse arrival times as a function of time is $\bm{\mu}(\bm{\theta})$, and $\bm{C}(\bm{\theta})$ is a covariance matrix that describes stochastic processes. 
Diagonal elements of $\bm{C}$, $\bm{\sigma}^2$, correspond to temporally-uncorrelated ``white'' noise. 
It is modelled as $\bm{\sigma}^2(\bm{e}_\text{f},\bm{e}_\text{q}) = \bm{\sigma}_\text{ToA}^2 \bm{e}_\text{f}^2 + \bm{e}_\text{q}^2$, where $\bm{\sigma}_\text{ToA}$ is the measurement uncertainty on the data provided by the initial timing model fit and $(\bm{e}_\text{f},\bm{e}_\text{q})$ are white noise model parameters\footnote{Also referred to as the error factor (EFAC) and the error in quadrature (EQUAD), respectively.} which are chosen to be separate for each telescope backend-receiver combination.
Off-diagonal elements of $\bm{C}$ describe temporal correlations.
For illustrative purposes, it is convenient to represent temporal correlations as components of $\bm{\mu}$ together with the timing model contributions using a reduced-rank approximation~\citep{LentatiAlexander2013,vanHaasterenVallisneri2014}:
\begin{equation}\label{eq:mu_expression}
    \bm{\mu} = \bm{F}\bm{a} + \bm{M}\bm{\epsilon} + \bm{U}\bm{j} + \bm{d}(t).
\end{equation}
Here, $\bm{F}$ are the Fourier sine and cosine basis function and $\bm{a}$ are Fourier amplitudes of red processes.
Timing model contributions are similarly modelled via the design matrix $\bm{M}$ and the coefficients $\bm{\epsilon}$ represent timing model parameters.
Whereas a type of white noise corresponding to timing delays $\bm{j}$ in units [s] that are the same for all ToAs obtained in the same observing epoch (``jitter'' noise) are modelled using the basis $\bm{U}$. 
Vector $\bm{d}(t)$ corresponds to other deterministic signals as a function of pulsar observation time. 
Deterministic signals include exponential dips associated with pulse shape changes \citep{GoncharovReardon2021}.
Let us denote $\bm{b} \equiv (\bm{a}, \bm{\epsilon}, \bm{j}) \in \bm{\theta}$.
Bayes theorem is used to obtain a posterior distribution of model parameters $\mathcal{P}(\bm{\theta} | \bm{\delta t})$ from the likelihood and the prior $\pi(\bm{\theta})$:
\begin{equation}
    \mathcal{P}(\bm{\theta} | \bm{\delta t}) = \frac{\mathcal{L}(\bm{\delta t} | \bm{\theta}) \pi(\bm{\theta})}{\mathcal{Z}},
\end{equation}
where $\mathcal{Z}$ is the integral of the numerator over $\bm{\theta}$, it is termed Bayesian evidence. 

\textit{Hierarchical inference} -- the approach of parameterising prior distributions -- is already a part of the standard PTA data analysis machinery. 
It is employed to reduce the parameter space and to make it physically-motivated. 
Instead of Fourier amplitudes $\bm{a}$ (two per frequency, one per sine term of $\bm{F}$ and one per cosine term) it is more convenient to measure power-law parameters $(A,\gamma)$ of the power spectral density of red processes:
\begin{equation}
\label{eqn:crn}
P(f|A,\gamma) = \frac{ A }{12 \pi ^2} \left(\frac{f}{f_{\rm yr}} \right)^{-\gamma}.
\end{equation}
Thus, the prior on $\bm{a}$ joins single pulsar likelihoods into a joint posterior:
\begin{equation}\label{eq:pta_posterior}
    \mathcal{P}(\bm{\theta}^{'}, \bm{\theta}^{''} | \bm{\delta t}) = \mathcal{Z}^{-1} \int \mathcal{L}(\bm{\delta t} | \bm{b},\bm{\theta}^{'}) \pi(\bm{b}|\bm{\theta}^{''}) \pi(\bm{\theta}^{''}) \pi(\bm{\theta}^{'}) d\bm{b},
\end{equation}
where $\bm{\theta}^{'}$ are parameters which are not included in the reduced-rank approximation in Equation~\ref{eq:mu_expression}, such that $\bm{\theta}=(\bm{\theta}^{'},\bm{b})$.
Values $\bm{\theta}^{''}$ are referred to as hyperparameters that include $(A,\gamma)$. 
The integral in Equation~\ref{eq:pta_posterior} is carried out analytically~\citep{LentatiAlexander2013,vanHaasterenVallisneri2014}. 
The integral represents \textit{marginalisation} over $\bm{b}$, and it is ultimately performed with $(\bm{F},\bm{M},\bm{U})$ and $\bm{b}$ modelled in $\bm{C}$.
For more details, please refer to~\citet{NG9_GWB}.

Full PTA analysis with stochastic processes that are correlated between pulsars, such as the gravitational wave background, is performed as follows.
To model the PSD of each pulsar-specific noise term, which may vary from pulsar to pulsar, we have a vector of power-law parameters $(\bm{A},\bm{\gamma})$, with a pair of noise (hyper-)parameters $(A,\gamma)$ per pulsar.
We also have a pair of $(A_\text{c},\gamma_\text{c})$ to model PSD of each ``common'' stochastic signal, \textit{i.e.}, applicable to all pulsars.
Using notations from Equation~\ref{eq:mu_expression}, an ``$\bm{a}$'' component of the term $\pi(\bm{b}|\bm{\theta}'')$ for arbitrary pulsars $(a,b)$ and frequencies $(i,j)$ becomes
\begin{equation}
\begin{split}
    \pi_{(a,i),(b,j)}(\bm{a}|A_a,\gamma_a,A_\text{c},\gamma_\text{c}) =  P_{ai}(f_i|A_a,\gamma_a) \delta_{ab} \delta_{ij}~+ \\ +~\Gamma_{ab} P_{i}(f_i|A_\text{c},\gamma_\text{c}) \delta_{ij},
\end{split}
\end{equation}
where $\Gamma_{ab}$ is the overlap reduction function such as the Hellings-Downs function.
It encodes pulsar-to-pulsar correlations.

\begin{table*}%[!htb]
\caption{\label{tab:hierarchicalmethods}Hierarchical Bayesian analysis methods for Pulsar Timing Arrays (PTAs). Parameters $\bm{\theta}$ of a model describing PTA data include parameters of noise and signals such as the gravitational wave background (GWB). Hyperparameters $\bm{\Lambda}$ determine the distribution of noise parameters in nature, $\pi(\bm{\theta}|\bm{\Lambda})$.}
\renewcommand{\arraystretch}{2.0}
\begin{tabularx}{\textwidth}{@{\extracolsep{\fill}} >{\hsize=0.6\hsize}L >{\hsize=0.4\hsize}C > {\hsize=0.4\hsize}C > {\hsize=0.6\hsize}C >{\hsize=1.2\hsize}C @{} }
\hline % {@{} L C C L @{} }

\hline \hline
Approach & Deliverable & Pulsar-wise parallelization & Modelling common signals in all pulsars & Primary scope \\ \hline
No modelling of ensemble noise properties, $\bm{\Lambda}=\varnothing$ & $\mathcal{P}(\bm{\theta}|\bm{\delta t})\bigg\rvert_{\bm{\Lambda}=\varnothing}$ & --- & Yes & Measurement of model parameters $\bm{\theta}$, subject to potential biases due to prior misspecification \\ 
\citet{GoncharovThrane2022}, $\bm{\Lambda}_\text{CP}=(\mu_{\lg A_\text{CP}},\sigma_{\lg A_\text{CP}})$~\citet{} & $\mathcal{P}(\bm{\Lambda}_\text{CP}|\bm{\delta t})$ & Yes & Via a prior in individual pulsars & Modelling of the common-spectrum process (CP, pulsar-to-pulsar autocorrelation of the GWB) \\ 
\citet{vanhaasteren2024} & $\mathcal{P}(\bm{\Lambda},\bm{\theta}|\bm{\delta t})$ & No & Yes & Inference of ensemble noise properties, resolving model misspecification \textit{on-the-fly} \\ 
Single-pulsar prior reweighting (Equation~\ref{eq:prior_reweighting}) & $\mathcal{P}(\bm{\Lambda}|\bm{\delta t})$  & Yes & Via a prior in individual pulsars & Inference of ensemble noise properties, resolving model misspecification \textit{in the subsequent analysis} \\ 
Marginalization over $\bm{\Lambda}$ (Equation~\ref{eq:h_marg}) & $\mathcal{P}(\bm{\theta}|\bm{\delta t})$ & No & Yes & Resolving model misspecification \textit{on-the-fly} \\ \hline \hline
\end{tabularx}
\end{table*}
% comparison table of methods: paper, purpose, allows to fit for HD simultaneously, common-spectrum signal

\subsection{\label{sec:hierarchical} Hierarchical inference of ensemble noise properties}

As part of the standard PTA analysis routine described in Section~\ref{sec:standard_methodology}, priors $\pi(\lg A,\gamma)$ are assumed to be uniform distributions, $\mathcal{U}$.
For example, $\pi(\lg A)=\mathcal{U}(-20,-12)$, $\pi(\gamma)=\mathcal{U}(0,7)$.
As we pointed out in the Introduction, previous studies suggest that this model may be incorrect. 
The solution is to propose alternative parameterised models $\pi(\bm{\theta}|\bm{\Lambda})$ which now depend on hyperparameters $\bm{\Lambda}$.
Therefore, generalising $(\lg A, \gamma) \in \bm{\theta}$, the posterior becomes
\begin{equation}\label{eq:h_likelihood}
    \mathcal{P}(\bm{\theta},\bm{\Lambda}|\bm{\delta t}) = \mathcal{Z}^{-1} \mathcal{L}(\bm{\delta t}|\bm{\theta}) \pi(\bm{\theta}|\bm{\Lambda}) \pi(\bm{\Lambda}).
\end{equation}
The approach of~\citet{vanhaasteren2024} is to directly evaluate $\mathcal{P}(\bm{\theta},\bm{\Lambda}|\bm{\delta t})$.
In this work, we propose computationally efficient marginalisation of this posterior over (a) $\bm{\theta}$, or (b) $\bm{\Lambda}$, as described below. 
A summary of the methods is provided in Table~\ref{tab:hierarchicalmethods}.

\subsubsection{\label{sec:hierarchical:reweigh} Marginalised posterior of hyperparameters}

To obtain $\mathcal{P}(\bm{\Lambda})$ marginalised over $\bm{\theta}$, we propose a solution based on prior reweighting, the application of importance sampling to priors. 
The calculation is done in two steps, both of which manifest one global fit to data. 
First, one obtains $\mathcal{P}(\bm{\theta}',\bm{\theta}''|\bm{\delta t})$ for every pulsar assuming a fixed prior on noise parameters $\pi(\bm{\theta}^k_i|{\varnothing})$. 
It is referred to  as the proposal distribution.
Here, $\varnothing$ means that hyperparameters are fixed as in the standard EPTA analysis~\citep{EPTA_DR2_NOISE}. 
Second, one uses the resulting posterior samples and evidence values ${\cal Z}_{\varnothing, i}$ to construct a likelihood marginalised over all parameters except $\bm{\Lambda}$, $\mathcal{L}(\bm{\delta t} | \bm{\Lambda})$.
It is shown in~\citet{ThraneTalbot2019} that the likelihood from Equation~\ref{eq:h_likelihood} can be written in the following form:
\begin{equation}\label{eq:prior_reweighting}
    \mathcal{L}(\bm{\delta t} | \bm{\Lambda}) = \prod_i^{N_\text{psrs}} \frac{{\cal Z}_{\varnothing, i}(\bm{\delta t}_i)}{n_i} \sum_k^{n_i} \frac{\pi(\bm{\theta}^k_i|\bm{\Lambda})}{\pi(\bm{\theta}^k_i|{\varnothing})}.
\end{equation}
In the above equation, $N_\text{psrs}$ is the number of pulsars in a PTA, $n_i$ is the number of posterior samples obtained for a given pulsar, $\theta^k_i$ is the $k$'th posterior sample for $i$'th pulsar. 
Ensemble noise properties of the EPTA presented in this work are obtained based on Equation~\ref{eq:prior_reweighting}. 
The limitation of this approach is that it is based on expressing the total PTA likelihood as the product of single-pulsar likelihoods. 
Therefore, the approach is blind to Hellings-Downs correlations or other inter-pulsar (angular) correlations in the timing data. 
Whereas the presence of temporal correlations with the same $(\lg A, \gamma)$ in all pulsars has to be imposed via a prior in individual pulsars. 
To model CP~\citep{NG12_GWB, GoncharovShannon2021, ChenCaballero2021} associated with the GWB~\citep{NG15_GWB, EPTA_DR2_GW, ReardonZic2023b} in our hierarchical likelihood defined by Equation~\ref{eq:prior_reweighting}, we impose an additional spin-noise-like term with $\gamma=13/3$ in individual pulsar likelihoods. 
The methodology and implications of this choice are discussed in Appendix \ref{sec:sim}.
Improved methods for separating the effect of gravitational wave background from pulsar-intrinsic work may be explored in future work. 

\subsubsection{\label{sec:hierarchical:marg} Marginalisation over hyperparameters}

The posterior from Equation~\ref{eq:h_likelihood} marginalised over $\bm{\Lambda}$ can be written in the following form:
\begin{equation}\label{eq:h_marg}
    \mathcal{P}(\bm{\theta}|\bm{\delta t}) =  \frac{\mathcal{L}(\bm{\delta t}|\bm{\theta}) \pi(\bm{\theta}|\varnothing)}{\mathcal{Z}} \times \frac{1}{n_\text{p}} \sum_{k}^{n_\text{p}} \frac{\pi(\bm{\theta}|\bm{\Lambda}_k)}{\pi(\bm{\theta}|\varnothing)},
\end{equation}
where $\bm{\Lambda}_k$ are $n_\text{p}$ samples from the prior. 
The reader may notice that the resulting posterior in Equation~\ref{eq:h_marg} is the product ($\times$) of the standard PTA posterior and the weight factor. 
A derivation of Equation~\ref{eq:h_marg} is provided in the Appendix. 
Unlike Equation~\ref{eq:prior_reweighting}, Equation ~\ref{eq:h_marg} allows to simultaneously model Hellings-Downs and other inter-pulsar correlations in the data while accounting for the uncertainty in pulsar noise priors. 
It is used more extensively in the companion paper~\citep{GoncharovSardana2024}, where the likelihood is multiplied by the weight factor from Equation~\ref{eq:h_marg}, and the rest of the analysis is performed in a standard way. 
Alternatively, the posterior $\mathcal{P}(\bm{\theta}|\bm{\delta t})$ can be reweighted into a $\bm{\Lambda}$-marginalised posterior using rejection sampling.

\section{Results}\label{sec:results}

In this work, we infer distributions for parameters governing pulsar spin noise, $\pi(\lg A_\text{SN},\gamma_\text{SN})$, and DM noise, $\pi(\lg A_\text{DM},\gamma_\text{DM})$, for EPTA DR2. 
EPTA DR2 contains CP associated with the GWB, such that $\gamma_\text{CP}$ is consistent with $13/3$~\citep{GoncharovSardana2024}. 
If we do not model this term, it may appear as a ``quasi-common'' red noise~\citep[as per the terminology of][]{GoncharovThrane2022}. 
In other words, as a separate cluster in the distribution of spin noise parameters. 
Because we are interested in knowing the distribution of \textit{pulsar-intrinsic} parameters, we include an additional red noise term with $\gamma_\text{c}=13/3$ to marginalise over a contribution of the common signal.
This is done during the analysis of single pulsar noise corresponding to the first step in Section~\ref{sec:hierarchical}. 
It corresponds to modelling common signal in our data via a prior as per Table~\ref{tab:hierarchicalmethods}. 
Because we are agnostic about $\lg A_\text{CP}$, the approach imposes a suboptimal uncertainty for measuring $(\lg A, \gamma)$ in pulsars where $\gamma$ is consistent with $13/3 \approx 4$. 
Furthermore, to separate contributions of other noise processes such as chromatic noise (including scattering variations) and band- or system-dependent noise, we adopt single-pulsar noise models and optimal numbers of Fourier frequency bins from the EPTA DR2 noise analysis~\citep{EPTA_DR2_NOISE}.

\subsection{\label{sec:results:unif} Uniform distribution of noise parameters}

\begin{figure*}%[!htb]
    \centering
    \begin{subfigure}[b]{0.48\textwidth}
        \includegraphics[width=\textwidth]{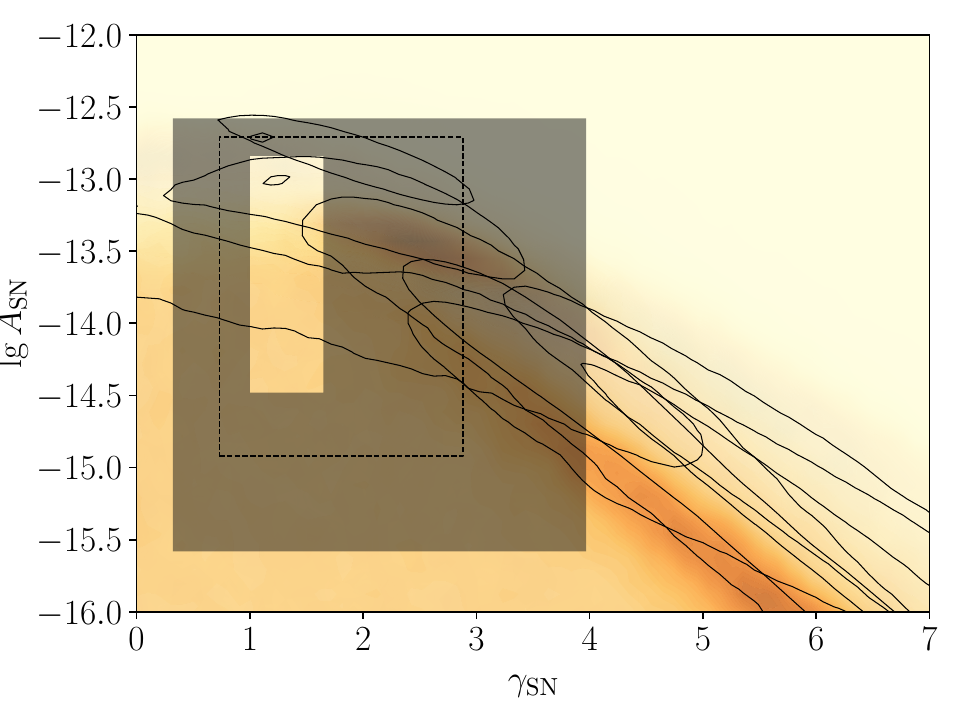}
        \caption{Spin noise (SN)}
        \label{fig:unif:sn}
    \end{subfigure}
    \begin{subfigure}[b]{0.48\textwidth}
        \includegraphics[width=\textwidth]{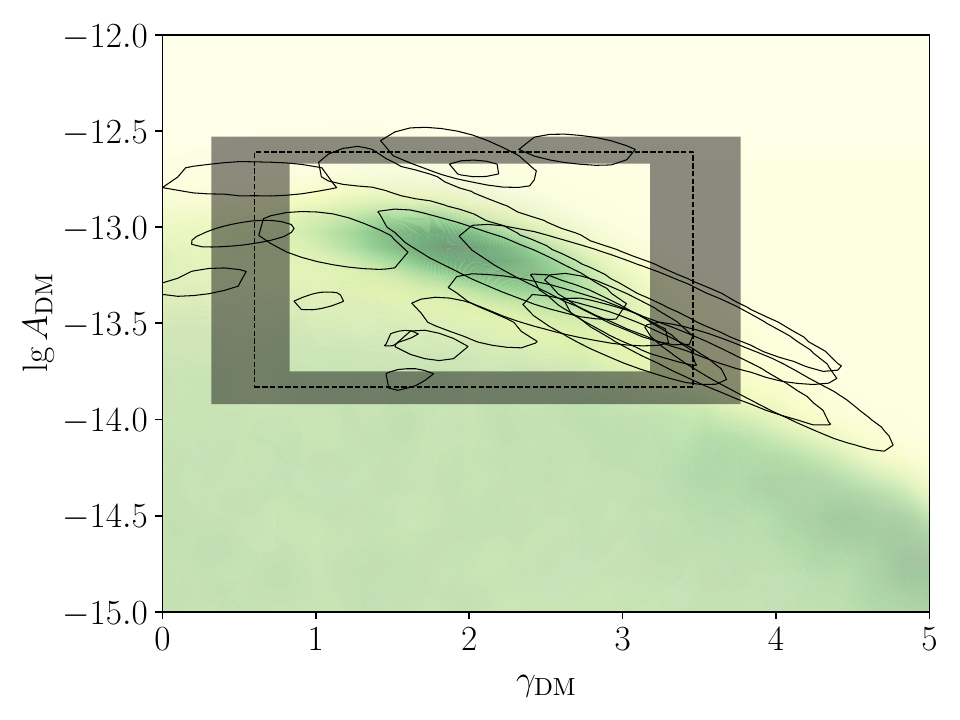}
        \caption{Dispersion measure (DM) variation noise}
        \label{fig:unif:dm}
    \end{subfigure}
    %\vspace{-1.2\baselineskip} % to shorten space
    \caption{The inferred form of the \textit{uniform} distribution of pulsar red noise parameters. 
    Black solid contours correspond to $1\sigma$ posterior levels of pulsar noise parameters when analysing single pulsar data, provided that the noise term is resolved in a pulsar based on model selection from \citet{EPTA_DR2_NOISE} (where inference is performed without the CP term, therefore our solid contours for SN may be different to those in the reference). 
    Colour corresponds to the posterior density of pulsars where noise terms were not resolved. 
    Our hierarchical analysis yields hyperparameters which determine how pulsar noise parameters are distributed. 
    The shaded rectangular areas correspond to $1\sigma$ credible levels for hyperparameters $(\bm{\min},\bm{\max})$ that govern limits of the uniform distribution based on our hierarchical analysis. 
    Hyperparameter values corresponding to the shaded area and the dashed line represent a row $\mathcal{U}(\bm{\theta} | \textbf{min},\textbf{max})$ in Table~\ref{tab:results}.
    } 
    \label{fig:unif}
\end{figure*}

We start with the case of \textit{uniform distributions} for noise parameters and we measure hyperparameters that govern uniform prior range.
For DM noise, these are $\text{min}(\lg{A}_\text{DM})$, $\text{max}(\lg{A}_\text{DM})$, $\text{min}(\gamma_\text{DM})$, $\text{max}(\gamma_\text{DM})$. 
We apply the same principle to spin noise (SN) parameters. 
Pulsar-specific measurements of $(\lg{A},\gamma)$ and the inferred boundaries of the uniform distribution for spin noise and DM noise parameters -- with measurement uncertainties -- are shown in Figure~\ref{fig:unif}. 
The first important observation is that the data rules out, with high credibility, the standard values of uniform prior boundaries. 
The data suggests that pulsar noise parameters are distributed in a more narrow range.
Posterior density $\mathcal{P}(\lg A, \gamma)$ from pulsars that do not have evidence for the respective noise term according to the~\citet{EPTA_DR2_NOISE} is shown in colour.
The data from these pulsars has also contributed to the measurement of hyperparameters. 

\subsection{\label{sec:results:norm} Normal distribution of noise parameters}

\begin{figure*}%[!htb]
    \centering
    \begin{subfigure}[b]{0.48\textwidth}
        \includegraphics[width=\textwidth]{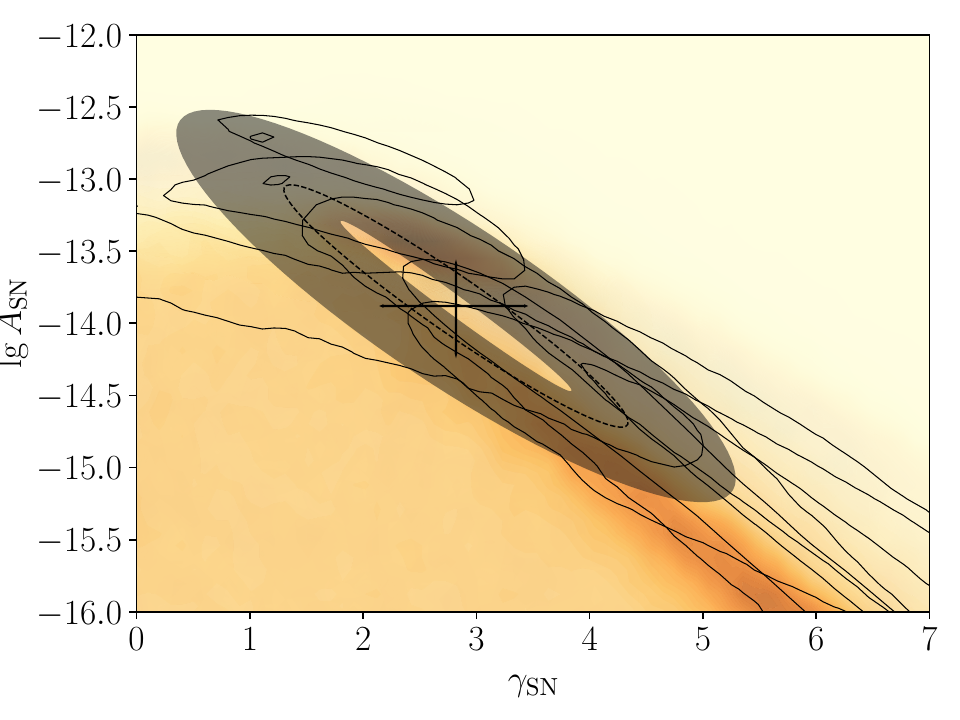}
        \caption{Spin noise (SN)}
        \label{fig:norm:sn}
    \end{subfigure}
    \begin{subfigure}[b]{0.48\textwidth}
        \includegraphics[width=\textwidth]{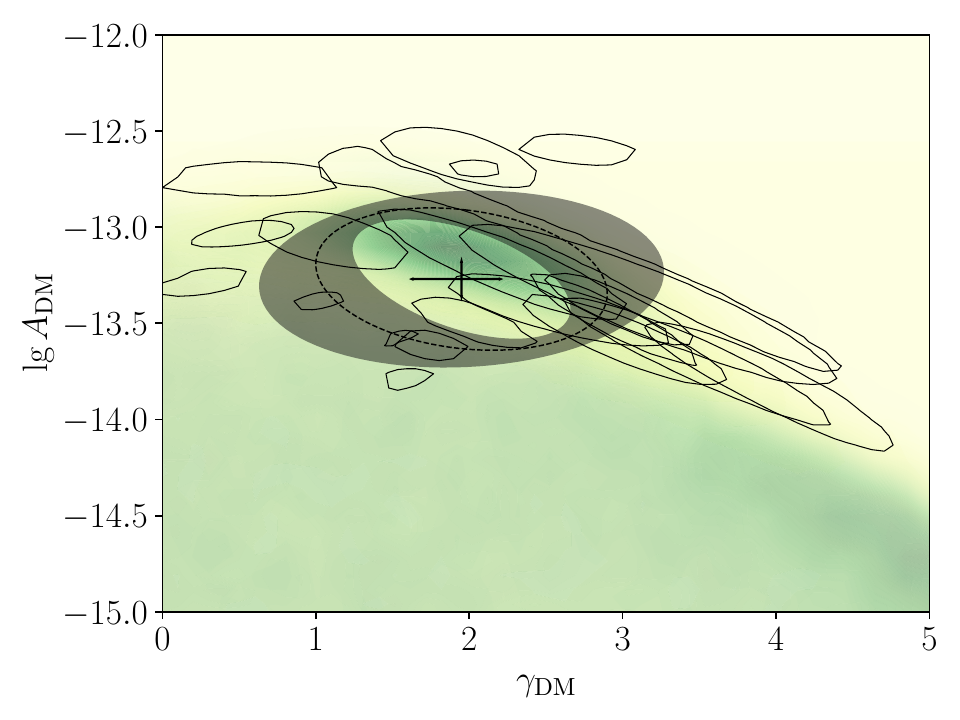}
        \caption{Dispersion measure (DM) variation noise}
        \label{fig:norm:dm}
    \end{subfigure}
    %\vspace{-1.2\baselineskip} % to shorten space
    \caption{The inferred form of the \textit{normal} distribution of pulsar red noise parameters. 
    The colour and the black solid contours represent measurements of noise parameters in individual pulsars, as explained in Figure~\ref{fig:unif}.
    Our hierarchical analysis yields hyperparameters which determine how pulsar noise parameters are distributed. 
    Shaded ellipses correspond to $1\sigma$ credible levels for hyperparameters $(\bm{\sigma},\rho)$ which govern the width of the distribution and the tilt, respectively. 
    Arrows correspond to $1\sigma$ credible levels for hyperparameters $\bm{\mu}$ which govern the position of the distribution. 
    Hyperparameter values corresponding to the shaded area and the dashed line represent a row $\mathcal{N}(\bm{\theta} |\bm{\mu},\bm{\sigma},\rho)$ in Table~\ref{tab:results}.
    } 
    \label{fig:norm}
\end{figure*}

The uniform prior model may remain a good approximation, but it is not the most natural choice. 
Therefore, we further discuss the possibility that pulsar noise parameters are \textit{normally distributed}. 
Thus, for spin noise, $\pi(\lg A_\text{SN},\gamma_\text{SN} | \bm{\Lambda}_\mathcal{N})$, and $\bm{\Lambda}_\mathcal{N}=(\mu_{\lg A_\text{SN}}$, $\sigma_{\lg A_\text{SN}}$, $\mu_{\gamma_\text{SN}}$, $\sigma_{\gamma_\text{SN}}$, $\rho_\text{SN})$.
The same model is applied to DM noise. 
Here, $\mu$ and $\sigma$ correspond to the mean and the standard deviation of the normal distribution of either $\lg A$ or $\gamma$, as subscripts indicate.
Parameter $\rho \in [-1,1]$ is the correlation coefficient between $\lg A$ and $\gamma$. 
It is important to note that we obtained our posterior samples from individual pulsars on step one from Section~\ref{sec:hierarchical} based on uniform priors.
So, for step two, there are no posterior samples to recycle from outside of these boundaries. 
Therefore, we truncate our normal distribution model and apply the same boundaries. 
The inferred boundaries of the truncated normal distribution for spin noise and DM noise parameters are shown in Figure~\ref{fig:norm}. 
The shaded area corresponds to an uncertainty in $(\bm{\sigma},\rho)$ and arrows separately show the uncertainty in $\bm{\mu}$. 
The uncertainty in $\rho$ is seen for DM noise as the difference between the tilt of the outer border of the shaded area compared to the inner border. 
Whereas for spin noise the value of $\rho \approx -1$ is strongly preferred. 

\subsection{\label{sec:results:mix} Distribution of noise parameters as a mixture model}

One may also be interested whether the distribution of pulsar noise parameters is more complex. 
For example, in Figure 1 in~\citet{GoncharovReardon2021}, one may notice the clustering of noise parameters in two areas.
First, around $\gamma \approx 1$.
Second, a cluster of pulsars around $\gamma=13/3$ of the gravitational wave background. 
A similar clustering is observed in the PTA noise analysis by the~\citet{NG15_NOISE}, where it is explained that pulsars with $\gamma \approx 13/3$ contribute to CP in full-PTA analysis. 
An analogous Figure is shown in~\citet{ReardonZic2023b}, although the presence of two over-densities is not obvious there.
Although we have taken steps to marginalize the contribution of the CP and uncover properties of pulsar-intrinsic noise, it is nevertheless useful to consider the possibility of a mixture model of a Gaussian distribution $\mathcal{N}$ and a Uniform distribution $\mathcal{U}$, where a broader Uniform distribution may fit potential outliers. 
The density of such a distribution is then $\nu\mathcal{N} + (1-\nu)\mathcal{U}$, where $\nu \in [0,1]$ is the contribution of the normal distribution.

\subsection{\label{sec:results:summary} Summary of the results}

As part of our measurements of hyperparameters, we have also obtained Bayesian evidence values $\mathcal{Z}$.
With this, we perform model selection to determine which of the models fit the observed distribution of pulsar noise parameters best. 
First, we find that the standard `static' uniform priors are disfavored compared to hierarchical uniform priors with natural log Bayes factors of $\gtrsim 100$. 
Therefore, the use of standard uniform priors is not recommended for PTA data analysis. 
The rest of the results of our hyperparameter estimation and hierarchical model selection are shown in Table~\ref{tab:results}. 

Some ensemble properties of spin noise (SN) are similar to DM noise, whereas some properties are different. 
For both SN and DM, for all models, the area of the parameterised prior is shrunk as much as possible to achieve the maximum posterior probability (given a fixed area, the narrower the distribution is - the higher it's maximum), but it also has to be sufficiently wide to be consistent with all the samples of noise amplitudes and spectral indices.
For spin noise, the normal distribution fits the data as well as the uniform distribution. 
However, the covariance between $\lg A_\text{SN}$ and $\gamma_\text{SN}$ plays an important role. 
In Figures~\ref{fig:unif:sn} and~\ref{fig:norm:sn}, one may notice a diagonal trend which results in finding a covariance parameter $\rho_\text{SN}\approx-1$. 
The model with a lack of covariance, $\rho_\text{SN}=0$ is disfavored with $\ln \mathcal{B}=6$. 
The covariance between $\lg A_\text{SN}$ and $\gamma_\text{SN}$ can be explained by noticing that it follows a line of equal noise power. 
In contrast, for DM noise, we do not find strong evidence for covariance between $\lg A_\text{DM}$ and $\gamma_\text{DM}$. 
With a lack of covariance, the uniform model is strongly preferred over the Gaussian model with a log Bayes factor of $6$. 
The mixture model is disfavored by our observations of both spin noise and DM noise, indicating the lack of outlying noise terms. 

Given the posterior biases found in simulated data studies in Appendix \ref{sec:sim}, we repeated all analyses from this section simultaneously with modelling the CP amplitude $\lg A_\text{CP}$ to be drawn from the truncated Normal distribution with hyperparameters $(\mu_{\lg A_\text{CP}},\sigma_{\lg A_\text{CP}})$. 
This is the approach that mitigated posterior biases in the simulations. 
SN and DM noise hyperparameters inferred this way are consistent with those in Table \ref{tab:results} at $1\sigma$ level, except for the mixture model's $\mu_{\gamma_\text{SN}}=1.44^{+0.62}_{-0.84}$. 
Because the mixture model is disfavored by the data and the rest of the results are consistent with Table \ref{tab:results}, we do not report the results obtained under this alternative model for other hyperparameters. 

\begin{table*}%[!htb]
\caption{\label{tab:results}Results of hyperparameter estimation and hierarchical model selection for ensemble pulsar noise properties of the European Pulsar Timing Array. DM noise and spin noise are parameterised by $\bm{\theta}=(\lg A,\gamma)$. We present the results for the following models of pulsar noise parameter distributions: a uniform distribution $\mathcal{U}$, a normal distribution $\mathcal{N}$ with and without covariance between $\lg A$ and $\gamma$, a mixture model of a normal and a uniform distribution. Columns $\mathcal{P}(\bm{\Lambda}|\bm{\delta t})$ contain \textit{maximum-aposteriori} hyperparameter values with $1\sigma$ credible levels. Columns $\ln \mathcal{B}$ contain log Bayes factors in favour of a model of interest against the uniform distribution model. The fact that they are negative means that the uniform model is the best fit. }
\renewcommand{\arraystretch}{2.0}
\begin{tabularx}{\textwidth}{@{\extracolsep{\fill}} |>{\hsize=0.6\hsize}L |>{\hsize=0.55\hsize}C >{\hsize=0.55\hsize}C > {\hsize=0.2\hsize}C |>{\hsize=0.55\hsize}C >{\hsize=0.55\hsize}C >{\hsize=0.2\hsize}C| @{} }
\hline % {@{} L C C L @{} }

\hline %\hline
\multirow{3}{*}{Model, $\pi(\bm{\theta}|\bm{\Lambda})$} & \multicolumn{3}{c|}{Achromatic ``spin'' noise (SN)} & \multicolumn{3}{c|}{Dispersion measure (DM) variation noise} \\ 
 & \multicolumn{3}{c|}{$\bm{\theta}=(A_\text{SN}, \gamma_\text{SN})$} & \multicolumn{3}{c|}{$\bm{\theta}=(A_\text{DM}, \gamma_\text{DM})$} \\ \cline{2-7} 
 & \multicolumn{2}{c}{$\mathcal{P}(\bm{\Lambda}_\text{SN}|\bm{\delta t})$} & $\ln \mathcal{B}_{\mathcal{U}_\text{SN}}$ & \multicolumn{2}{c}{$\mathcal{P}(\bm{\Lambda}_\text{DM}|\bm{\delta t})$} & $\ln \mathcal{B}_{\mathcal{U}_\text{DM}}$ \\ \hline \hline

% UNIFORM
\multirow{4}{*}{\makecell{$\mathcal{U}(\bm{\theta} | \textbf{min},\textbf{max})$\\\\Shown in Figure~\ref{fig:unif}}} & $\min(\lg A_\text{SN})$ & $-14.91^{+0.43}_{-0.66}$ & \multirow{4}{*}{-} & $\min(\lg A_\text{DM})$ & $-13.83^{+0.08}_{-0.09}$ & \multirow{4}{*}{-} \\ 
 & $\max(\lg A_\text{SN})$ & $-12.71^{+0.13}_{-0.13}$ &  & $\max(\lg A_\text{DM})$ & $-12.61^{+0.08}_{-0.06}$ &  \\ 
 & $\min(\gamma_\text{SN})$ & $ 0.73^{+0.27}_{-0.41}$ &  & $\min(\gamma_\text{DM})$ & $ 0.60^{+0.23}_{-0.28}$ &  \\ 
 & $\max(\gamma_\text{SN})$ & $ 2.88^{+1.09}_{-1.23}$ &  & $\max(\gamma_\text{DM})$ & $ 3.46^{+0.31}_{-0.31}$ &  \\ \hline

% BIVARIATE TRUNCATED GAUSSIAN WITH COVARIANCE
\multirow{5}{*}{\makecell{$\mathcal{N}(\bm{\theta} |\bm{\mu},\bm{\sigma},\rho)$\\\\Shown in Figure~\ref{fig:norm}}} & $\mu_{\lg A_\text{SN}}$ & $-13.88^{+0.29}_{-0.33}$ & \multirow{5}{*}{$-0.6$} & $\mu_{\lg A_\text{DM}}$ & $-13.27^{+0.09}_{-0.09}$ & \multirow{5}{*}{$-6.0$} \\ % -0.6242287478, -5.9881408904
 & $\sigma_{\lg A_\text{SN}}$ & $ 0.84^{+0.52}_{-0.25}$ &  & $\sigma_{\lg A_\text{DM}}$ & $ 0.37^{+0.09}_{-0.06}$ &  \\ 
 & $\mu_{\gamma_\text{SN}}$ & $ 2.82^{+0.61}_{-0.65}$ &  & $\mu_{\gamma_\text{DM}}$ & $ 1.95^{+0.25}_{-0.32}$ &  \\ 
 & $\sigma_{\gamma_\text{SN}}$ & $ 1.52^{+0.95}_{-0.50}$ &  & $\sigma_{\gamma_\text{DM}}$ & $ 0.95^{+0.37}_{-0.24}$ &  \\ 
  & $\rho_\text{SN}$ & $ -0.96^{+0.08}_{-0.03}$ &  & $\rho_\text{DM}$ & $ -0.21^{+0.29}_{-0.29}$ &  \\ \hline

% BIVARIATE TRUNCATED GAUSSIAN WITHOUT COVARIANCE
\multirow{4}{*}{$\mathcal{N}(\bm{\theta} |\bm{\mu},\bm{\sigma},\rho=0)$} & $\mu_{\lg A_\text{SN}}$ & $-13.71^{+0.18}_{-0.26}$ & \multirow{4}{*}{$-6.5$} & $\mu_{\lg A_\text{DM}}$ & $-13.27^{+0.08}_{-0.08}$ & \multirow{4}{*}{$-6.6$} \\ % -6.4847870644, -6.6000044715
 & $\sigma_{\lg A_\text{SN}}$ & $ 0.54^{+0.21}_{-0.15}$ &  & $\sigma_{\lg A_\text{DM}}$ & $ 0.35^{+0.07}_{-0.05}$ &  \\ 
 & $\mu_{\gamma_\text{SN}}$ & $ 1.43^{+0.69}_{-0.16}$ &  & $\mu_{\gamma_\text{DM}}$ & $ 2.01^{+0.19}_{-0.19}$ &  \\ 
 & $\sigma_{\gamma_\text{SN}}$ & $ 0.34^{+0.51}_{-0.26}$ &  & $\sigma_{\gamma_\text{DM}}$ & $ 0.77^{+0.18}_{-0.15}$ &  \\ \hline

% MIXTURE BIVARIATE TRUNCATED GAUSSIAN + UNIFORM
\multirow{6}{*}{\makecell{$\nu\mathcal{N} + (1-\nu)\mathcal{U}$,\\\\$0 \leq \nu \leq 1$}} & $\mu_{\lg A_\text{SN}}$ & $-13.86^{+0.28}_{-0.29}$ & \multirow{6}{*}{$-3.7$} & $\mu_{\lg A_\text{DM}}$ & $-13.28^{+0.09}_{-0.08}$ & \multirow{6}{*}{$-9.1$} \\ % -3.7395255527, -9.1248939111
 & $\sigma_{\lg A_\text{SN}}$ & $ 0.78^{+0.43}_{-0.22}$ &  & $\sigma_{\lg A_\text{DM}}$ & $ 0.37^{+0.08}_{-0.06}$ &  \\ 
 & $\mu_{\gamma_\text{SN}}$ & $ 2.87^{+0.55}_{-0.67}$ &  & $\mu_{\gamma_\text{DM}}$ & $ 1.94^{+0.26}_{-0.34}$ &  \\ 
 & $\sigma_{\gamma_\text{SN}}$ & $ 1.48^{+0.86}_{-0.44}$ &  & $\sigma_{\gamma_\text{DM}}$ & $ 0.97^{+0.40}_{-0.24}$ &  \\ 
  & $\rho_\text{SN}$ & $-0.97^{+0.07}_{-0.02}$ &  & $\rho_\text{DM}$ & $-0.23^{+0.29}_{-0.27}$ &  \\ 
    & $\nu_\text{SN}$ & $ 0.95^{+0.03}_{-0.07}$ &  & $\nu_\text{DM}$ & $ 0.96^{+0.03}_{-0.05}$ &  \\ \hline
%$\mathcal{U}$ & - & Yes & - & - \\  \hline %\hline
\end{tabularx}

\end{table*}

\section{Caveats of the circular analysis}\label{sec:doubledipping} 

Circular analysis or ``double dipping'' refers to the usage of data to inform on the model which is then applied to analyse the same data. 
The methodology we propose in our study is not a circular analysis, it represents one global fit to PTA data. 
Despite the importance sampling involves performing parameter estimation on the same data twice, the final result is ultimately independent of the proposal distribution obtained at the first step. 
Namely, the target distribution is independent of the proposal distribution. 
However, because circular analysis has been extensively discussed in the PTA community, in this section we take an opportunity to cover this subject from the hierarchical inference standpoint. 

Overall, circular analyses may lead to the underestimation of measurement uncertainty and systematic errors. 
However, some PTA analyses in the past have involved a circular analysis, as pointed out by~\citet{vanHaasteren2025}.
Namely, pulsar spin noise and DM noise terms were included (not included) in the full-PTA analysis based on (a lack of) evidence for these terms in single-pulsar noise model selection. 
\citet{vanHaasteren2025} argues against this approach. 
Indeed, double dipping is not a proper statistical treatment, it should be abandoned in favour of the proper model selection and model averaging performed in one global fit, as recommended by the author.
However, to give a fair overview of the problem, we also show that circular analysis does not always yield incorrect conclusions in analyses of PTA data. 

Before we discuss the application of hierarchical inference in the subsections below, we would like to point out that it is not clear from~\citet{vanHaasteren2025} that the aforementioned approach of selecting spin noise terms for subsequent full-PTA analyses has led to underestimation of measurement uncertainties reported by PTAs. 
For model averaging, the prior odds between the existence or absence of CP are proposed to be 50\%-50\% in~\citet{vanHaasteren2025}.
For the case of including a red noise term to every pulsar, the prior odds will be arbitrary and vary from pulsar to pulsar which is clear from the equations in the Appendix in~\citet{vanHaasteren2025}.
However, data informs on these odds, so they should become a model (hyper-)parameter to avoid prior misspecification, the same problem discussed in this work.
These parameterised odds will act as model selection. 
In the limit where there is no measurable red noise in pulsars and the data informs on it sufficiently well, this will lead to the elimination of the contribution of one of the models. 
Because different noise models lead to different measurement uncertainties for $(\lg A_\text{CP},\gamma_\text{CP})$, it is possible that the uncertainty introduced by enforcing red noise to every pulsar is suboptimal. 

\subsection{\label{sec:doubledipping:comparison} A comparison between circular analysis, incorrect priors, and a proper analysis}

\begin{figure}%[!htb]
    \centering
    \begin{subfigure}[b]{0.35\textwidth}
        \includegraphics[width=\textwidth]{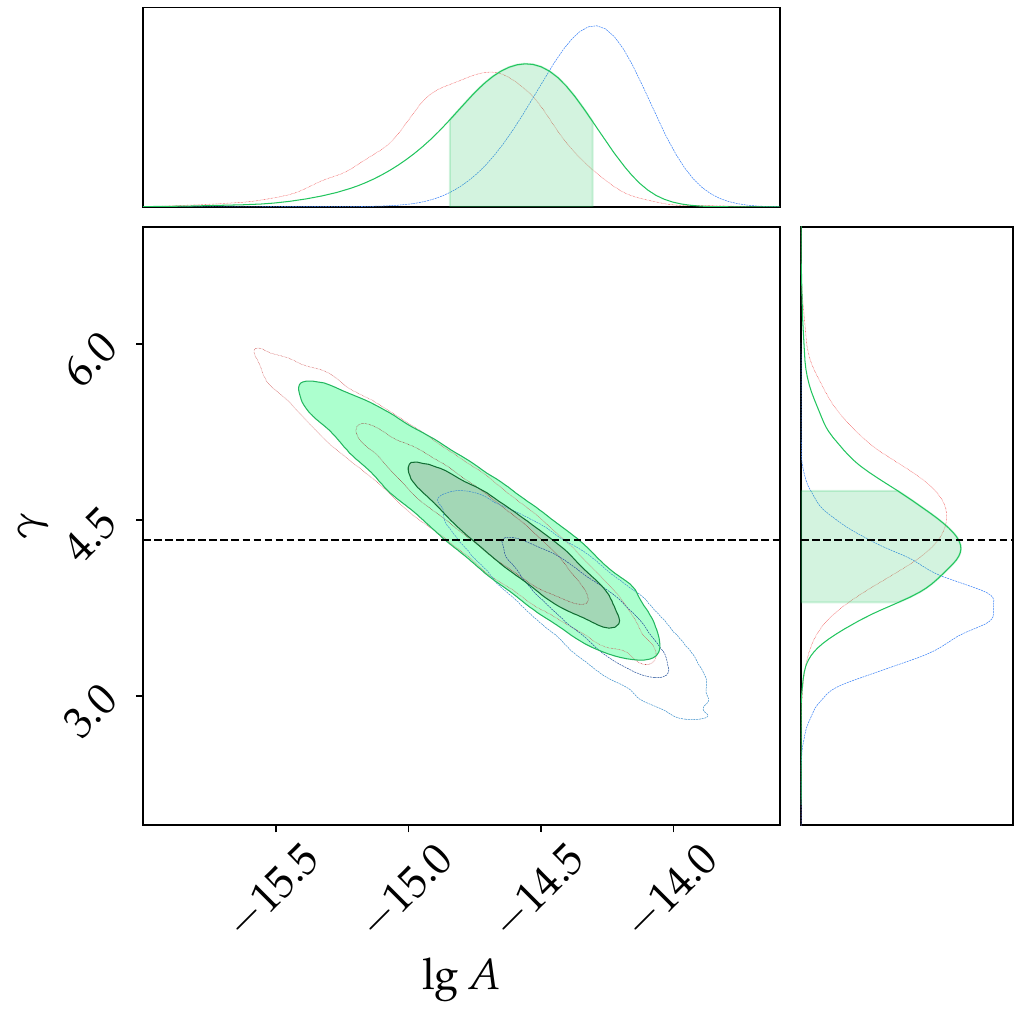}
        %\caption{1}
        %\label{fig:doubledip:comparison}
    \end{subfigure}
    %\vspace{-1.2\baselineskip} % to shorten space
    \caption{A comparison between circular analysis (red dotted contours), incorrect broad uniform priors (blue dashed contours), and a proper hierarchical analysis (green contours) of the common-spectrum process (CP) associated with the gravitational wave background (GWB).
    } 
    \label{fig:doubledip}
\end{figure}

One example of circular analysis in the context of our hierarchical inference is finding best-fit $\bm{\Lambda}$ and using it for the gravitational wave search. 
It also leads to resolving prior misspecification but potentially at the cost of a reduced measurement uncertainty for $(\lg A, \gamma)$ of the gravitational wave background. 
The reduction may come thanks to simply ignoring a part of the intrinsic measurement uncertainty for $\bm{\Lambda}$, so a regular noise fluctuation is more likely to render our estimate of $(\lg A, \gamma)$ to be inconsistent with the true value. 
In Figure~\ref{fig:doubledip}, we show three measurements of $(\lg A, \gamma)$ of the CP in the EPTA data. 
Blue dashed contours correspond to a posterior obtained with the standard uniform spin noise priors, $\pi(\bm{\theta}|\bm{\Lambda})=\mathcal{U}(\bm{\theta}|\varnothing)$. 
Red dotted contours correspond to $\pi(\bm{\theta}|\bm{\Lambda})=\mathcal{N}(\bm{\theta} |\bm{\mu}_0,\bm{\sigma}_0,\rho=0)$, where $(\bm{\mu}_0,\bm{\sigma}_0)$ are the values obtained for this model from Table \ref{tab:results}. 
Therefore, the red dotted line corresponds to a circular analysis. 
Filled green contours correspond to a proper (global-fit) measurement obtained based on Equation \ref{eq:h_marg}, it is discussed in more detail in~\citet{GoncharovSardana2024}. 
The measurement uncertainty of $(\lg A, \gamma)$ obtained with circular analysis matches that obtained with a full hierarchical analysis. 
It is even $5$-$8$\% larger, suggesting that in the current setting circular analysis overestimates the intrinsic measurement uncertainty. 
Note, however, that the circular analysis shifts the \textit{maximum-aposteriori} value of $(\lg A, \gamma)$, so the use of it is still not recommended. 

\subsection{\label{sec:doubledipping:quasicommon} A toy example: circular analysis of CP}

\begin{figure*}%[!htb]
    \centering
    \begin{subfigure}[b]{0.33\textwidth}
        \includegraphics[width=\textwidth]{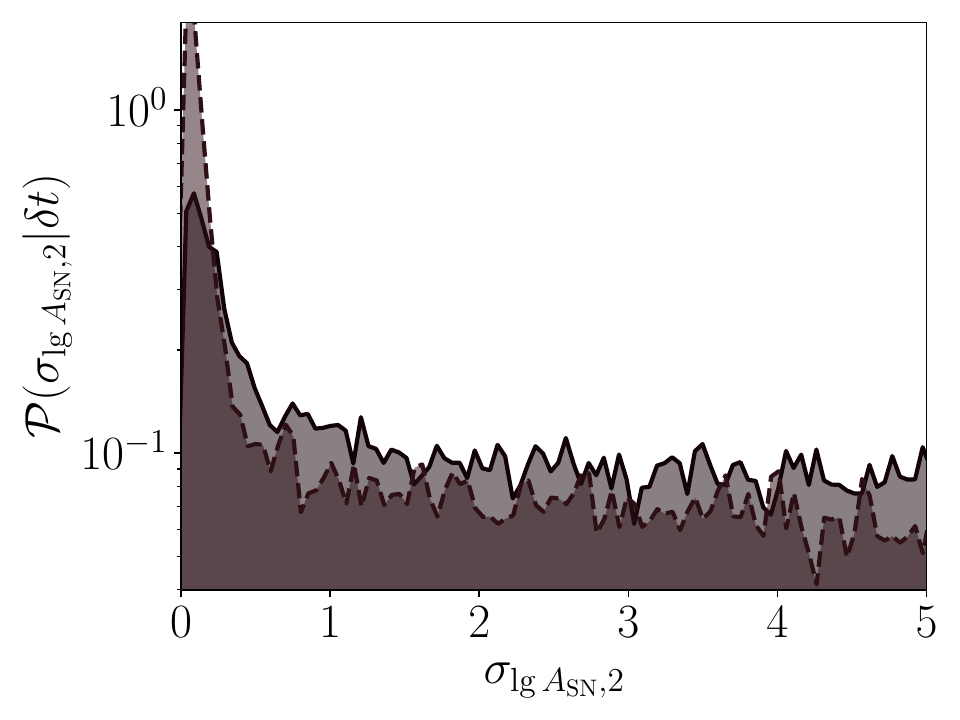}
        %\caption{1}
        \label{fig:quasicommon:sig_lgA2}
    \end{subfigure}
    \begin{subfigure}[b]{0.33\textwidth}
        \includegraphics[width=\textwidth]{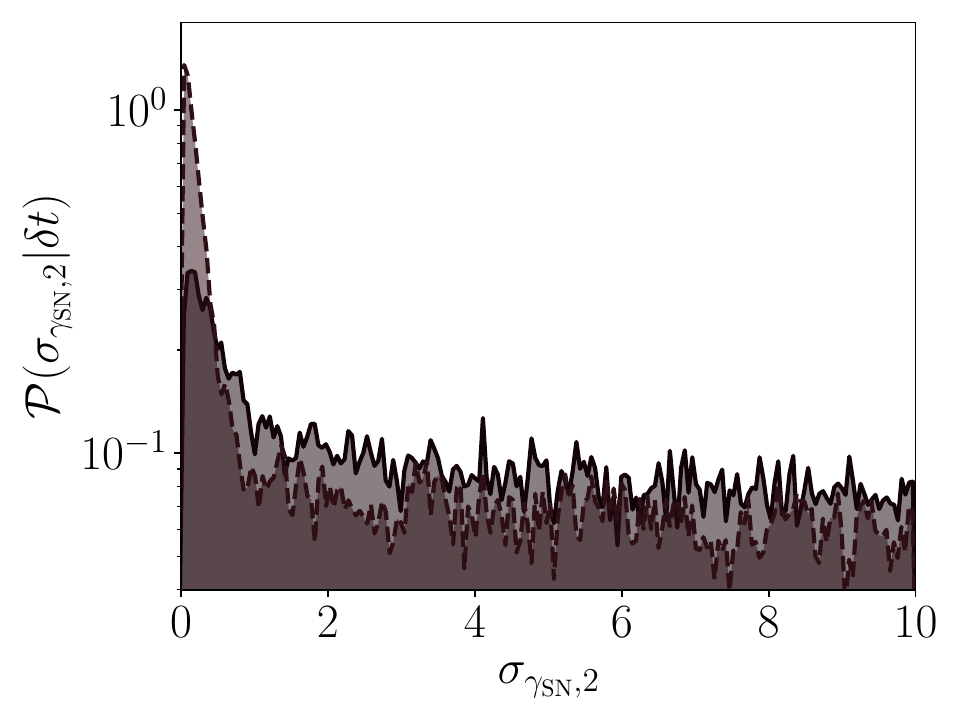}
        %\caption{2}
        \label{fig:quasicommon:sig_gam2}
    \end{subfigure}
    \begin{subfigure}[b]{0.33\textwidth}
        \includegraphics[width=\textwidth]{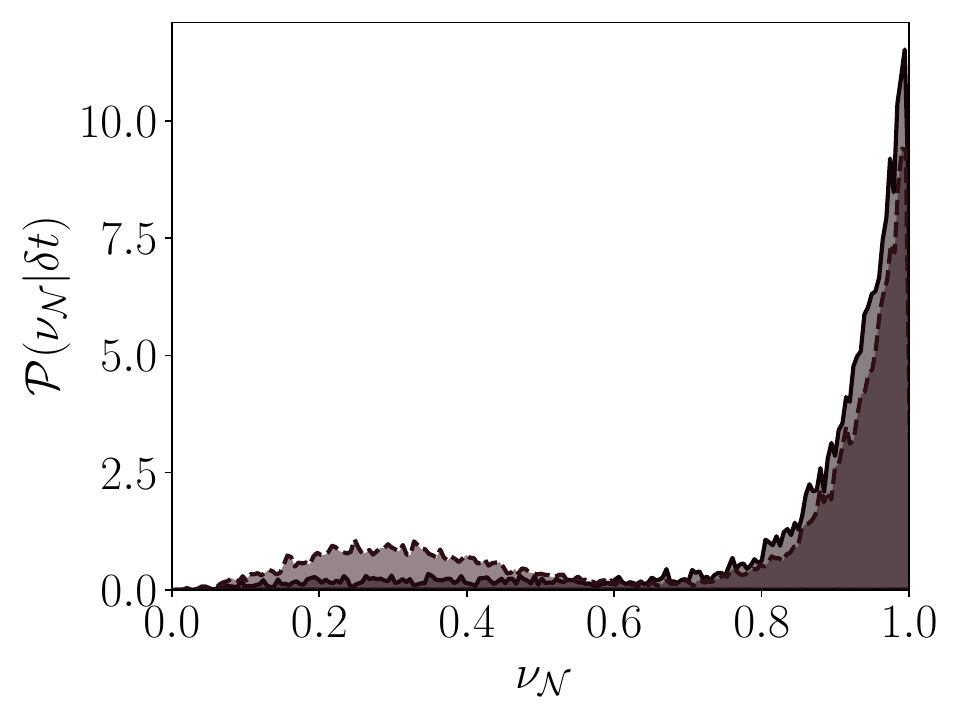}
        %\caption{2}
        \label{fig:quasicommon:nu}
    \end{subfigure}
    \vspace{-2.2\baselineskip} % to shorten space
    \caption{Modelling of the common-spectrum process (CP) as a narrow normally-distributed cluster of pulsar noise using data-informed hyperpriors $\pi(\bm{\Lambda})$. 
    We assume that pulsar noise parameters $\lg A$ and $\gamma$ are distributed as a mixture model of two Gaussians, where hyperparameters of the second Gaussian are chosen to match the measurement uncertainty on the common spectrum process reported by the~\citet{EPTA_DR2_GW}. 
    Parameters $\sigma$ with respective indices correspond to standard deviations of the distribution that governs $\lg A$ and $\gamma$. 
    Consistency of $\sigma$ with zero has first been shown in~\citet{GoncharovThrane2022} using a more rigorous although less flexible methodology. 
    Parameter $\nu_{\mathcal{N}}$ corresponds to the fraction of the first, wider normal distribution in a model. Solid lines correspond to including the second $\gamma=13/3$ red noise term to pulsars to account for the presence of the CP, dashed lines correspond to not including it. 
    As expected, the contribution of the CP becomes more visible in the latter case at $\sigma_{\lg A, \gamma}=0$ and $\nu_\mathcal{N} \approx 0.3$. 
    }
    \label{fig:quasicommon}
\end{figure*}

Keeping in mind the caveats above, for the test purposes we also perform the estimation of hyperparameters $\bm{\Lambda}$ for the mixture model of two Gaussian distributions where hyper-priors on the second component of the mixture correspond to EPTA measurement of the CP parameters~\cite{EPTA_DR2_GW}. 
This exercise remains useful to explore the interplay between pulsar-intrinsic noise and the CP and to understand the performance of the model. 
For the two-Gaussian mixture model, $\pi(\lg A_\text{SN},\gamma_\text{SN} | \bm{\Lambda}_{2\mathcal{N}})$, and $\bm{\Lambda}_{2\mathcal{N}}=(\mu_{\lg A_\text{SN}}^{1,2}$, $\sigma_{\lg A_\text{SN}}^{1,2}$, $\mu_{\gamma_\text{SN}}^{1,2}$, $\sigma_{\gamma_\text{SN}}^{1,2}$, $\rho_\text{SN}^{1,2}$, and $\nu_\mathcal{N})$.
Indices $(1,2)$ refer to each of the two components in a mixture, while $\nu$ corresponds to a fraction of the first mixture component in the total prior probability. 
For the second component to correspond to the CP in the EPTA data, we impose a Gaussian hyperprior determined by the measurement uncertainty on $(\lg A,\gamma)$ reported by the EPTA for our data. 
We find that the fraction of the second normal component is mostly consistent with one and inconsistent with zero. 
However, hyperparameters of the second mixture component -- mean and (co-)variance -- are well-constrained. 
Interestingly, we also find a marginal excess posterior density at $(\sigma_{\lg A_\text{SN}}^{2},\sigma_{\gamma_\text{SN}}^{2})=(0,0)$. 
This result is related to~\citet{GoncharovThrane2022}'s measurement of $\sigma_{\lg A}$ to be consistent with zero.
When removing the CP term with $\gamma=13/3$ from our pulsar-intrinsic noise model, the posterior density increases, as expected.
This is shown in Figure~\ref{fig:quasicommon}.

\section{Conclusions}\label{sec:conclusion} 

We performed inference of ensemble properties of spin noise and DM noise in the 25-year version of the second data release of the European Pulsar Timing Array (EPTA). 
Overall, we find that the standard uniform priors used in the previous analysis are too wide and not representative of the observations. 
However, a parameterised uniform prior distribution is a good fit. 
We, therefore, recommend correctly accounting for ensemble pulsar noise properties using Equation~\ref{eq:h_marg} in future full-PTA analyses to avoid systematic errors (parameter estimation and model selection biases). 
Our approach of prior reweighting for inferring noise hyperparameters based on Equation \ref{eq:prior_reweighting} is well-suited for understanding broad ensemble noise properties for simulating data and predicting the PTA noise budget. 

\section*{Acknowledgements}

We thank Rutger van Haasteren for insightful discussions about hierarchical inference and circular analysis. 
Some of our calculations were carried out using the OzSTAR Australian national facility (high-performance computing) at Swinburne University of Technology. 
%We also thank Alberto Sesana for helpful comments on the manuscript. 

%%%%%%%%%%%%%%%%%%%%%%%%%%%%%%%%%%%%%%%%%%%%%%%%%%
\section*{Data Availability}

The code to reproduce the results of this analysis and the analysis by~\citet{GoncharovThrane2022} is available at \href{https://github.com/bvgoncharov/pta_priors}{github.com/bvgoncharov/pta\_priors}. 
The data used for our study, EPTA DR2 by the~\citep{EPTA_DR2_TIMING}, is available at \href{zenodo}{zenodo}.
Other data may be provided by the corresponding author upon request. 
The PTA likelihood is incorporated in \textsc{enterprise}~\citep{enterprise} and posterior sampling is performed using \textsc{ptmcmcsampler}~\citep{ptmcmcsampler} and \textsc{dynesty}~\citep{Speagle2020}.

%%%%%%%%%%%%%%%%%%%% REFERENCES %%%%%%%%%%%%%%%%%%

% The best way to enter references is to use BibTeX:

\bibliographystyle{mnras}
\bibliography{mybib,collab_papers,soft} % if your bibtex file is called example.bib

%%%%%%%%%%%%%%%%%%%%%%%%%%%%%%%%%%%%%%%%%%%%%%%%%%

%%%%%%%%%%%%%%%%% APPENDICES %%%%%%%%%%%%%%%%%%%%%

\appendix 

\section{Tests with the simulated data}\label{sec:sim}

We test our inference of hyperparameters based on Section \ref{sec:hierarchical:reweigh} using two simulated data sets. 
Both simulations are based on $25$ pulsars, timed over $15$ years with the $100$~ns precision. 
The cadence is empirically chosen to visually represent irregular PTA observations. 
It is based on the Pareto distribution with the shape value of $0.6$, scaled to the PTA observation time, with the maximum observation gap of $0.6$ of the time span. 
Simulations contain spin noise and the common-spectrum process (CP). 
Spin noise is drawn from the truncated Normal distribution, with hyperparameters given by the first five columns in Table~\ref{tab:sim}. 
CP represents an additional red noise term in every pulsar, with the same amplitude and spectral index listed in the last two columns in Table~\ref{tab:sim}. 
Therefore, CP shows the expected spectral properties of the GWB from supermassive black hole binaries.

\begin{table}%[!htb]
\caption{\label{tab:sim}Parameters of our two simulations.}
\renewcommand{\arraystretch}{2.0}
\begin{tabularx}{\columnwidth}{@{\extracolsep{\fill}} >{\hsize=0.1\hsize}L >{\hsize=0.1\hsize}C >{\hsize=0.1\hsize}C >{\hsize=0.1\hsize}C >{\hsize=0.1\hsize}C >{\hsize=0.1\hsize}C >{\hsize=0.1\hsize}C @{} }
\hline % {@{} L C C L @{} }

\hline \hline
$\mu_{\lg A}$ & $\sigma_{\lg A}$ & $\mu_\gamma$ & $\sigma_\gamma$ & $\rho$ & $\lg A_\text{CP}$ & $\lg \gamma_\text{CP}$ \\ \hline
$-14.5$ & $1.5$ & $4.0$ & $1.5$ & \multirow{2}{*}{$0.0$} & \multirow{2}{*}{$-15.0$} &  \multirow{2}{*}{$13/3$}  \\ 
$-13.8$ & $0.7$ & $2.0$ & $1.0$ & &  &  \\ \hline \hline
\end{tabularx}
\end{table}

The simulations are designed to test our ability to correctly model the distribution of spin noise parameters given the presence of CP. 
In the first simulated data set, hyperparameters are chosen such that spin noise amplitudes and spectral indices significantly overlap with those of the CP. 
Because CP and spin noise are indistinguishable in the data of single pulsars, the proposal posteriors we construct from such data may show significant differences from the target posteriors we would get in a global fit to all pulsar data. 
As described in Section \ref{sec:hierarchical:reweigh}, we fix the spectral index of CP to $13/3$. 
Thus, in pulsars where $\gamma_\text{SN}$ is close to $13/3$, proposal posteriors on $A_\text{CP}$ peak at values of $A_\text{SN}$, with only the tails extending to the true value of $A_\text{CP}$. 
These biases are mitigated when obtaining target posteriors in the global fit, but the residual biases may remain. 
In the second simulated data, the overlap between CP and spin noise is less significant. 
Therefore, the first simulation is considered to be the most difficult case. 

\begin{figure*}%[!htb]
    \centering
    %\begin{subfigure}[b]{0.33\textwidth}
        \includegraphics[width=\textwidth]{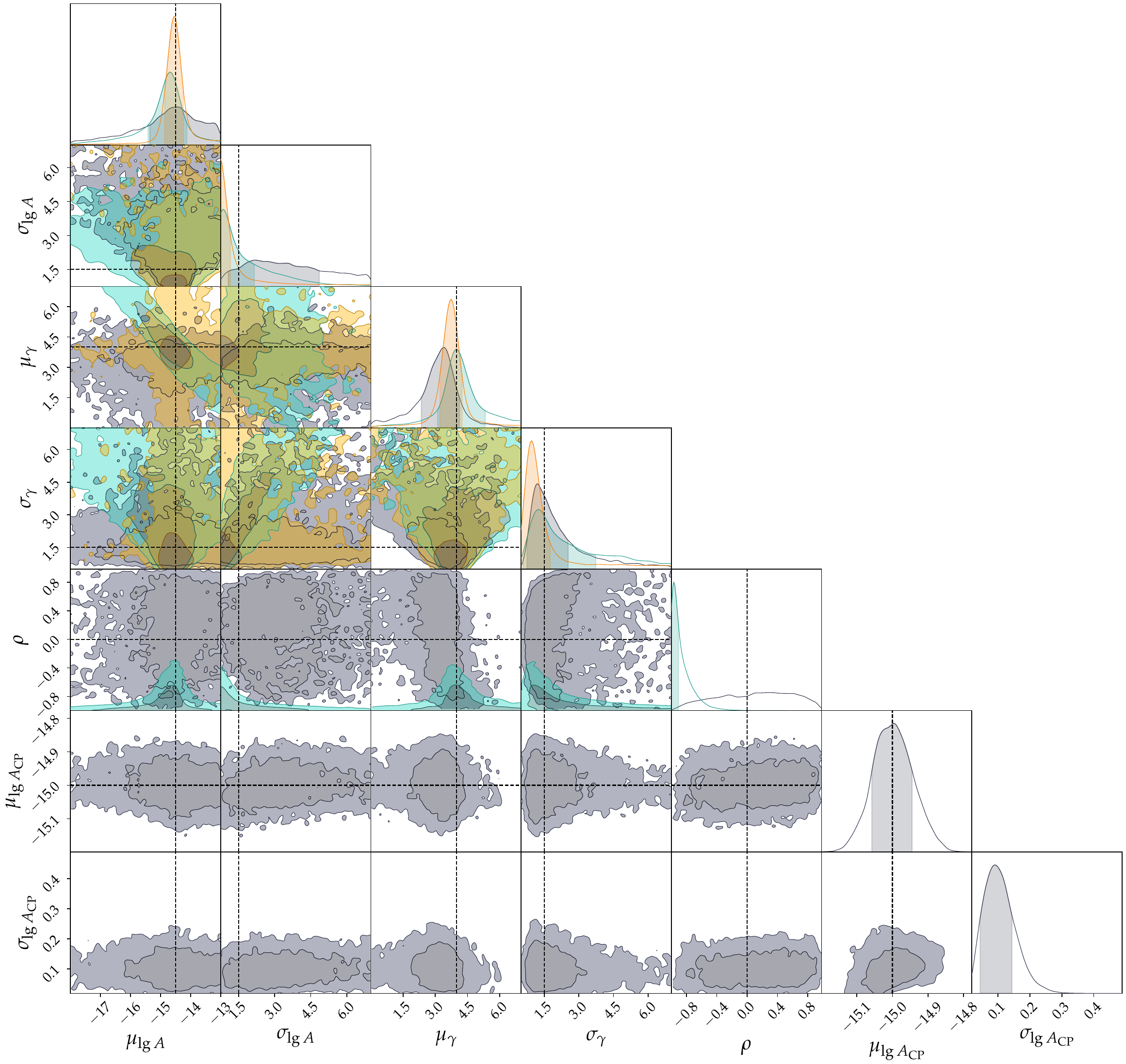}
        %\caption{1}
        \label{fig:quasicommon:sig_lgA2}
    %\end{subfigure}
    %\vspace{-2.2\baselineskip} % to shorten space
    \caption{Posteriors on the inferred hyperparameters in the first simulation. Dashed vertical lines correspond to simulated values in Table \ref{tab:sim}. Yellow contours corresponds to $\rho=0$, green contours correspond to $\rho$ as a free parameter, and black contours correspond to additionally modelling the distribution of CP amplitudes in pulsars as being drawn from the truncated Normal distribution with mean $\mu_{\lg A_\text{CP}}$ and the standard deviation $\sigma_{\lg A_\text{CP}}$. Shaded areas correspond to $1\sigma$ ($1\sigma$ and $2\sigma$) credible levels in fully-marginalised posteriors (two-dimensional posteriors).}
    \label{fig:sim1}
\end{figure*}

\begin{figure*}%[!htb]
    \centering
    %\begin{subfigure}[b]{0.33\textwidth}
        \includegraphics[width=\textwidth]{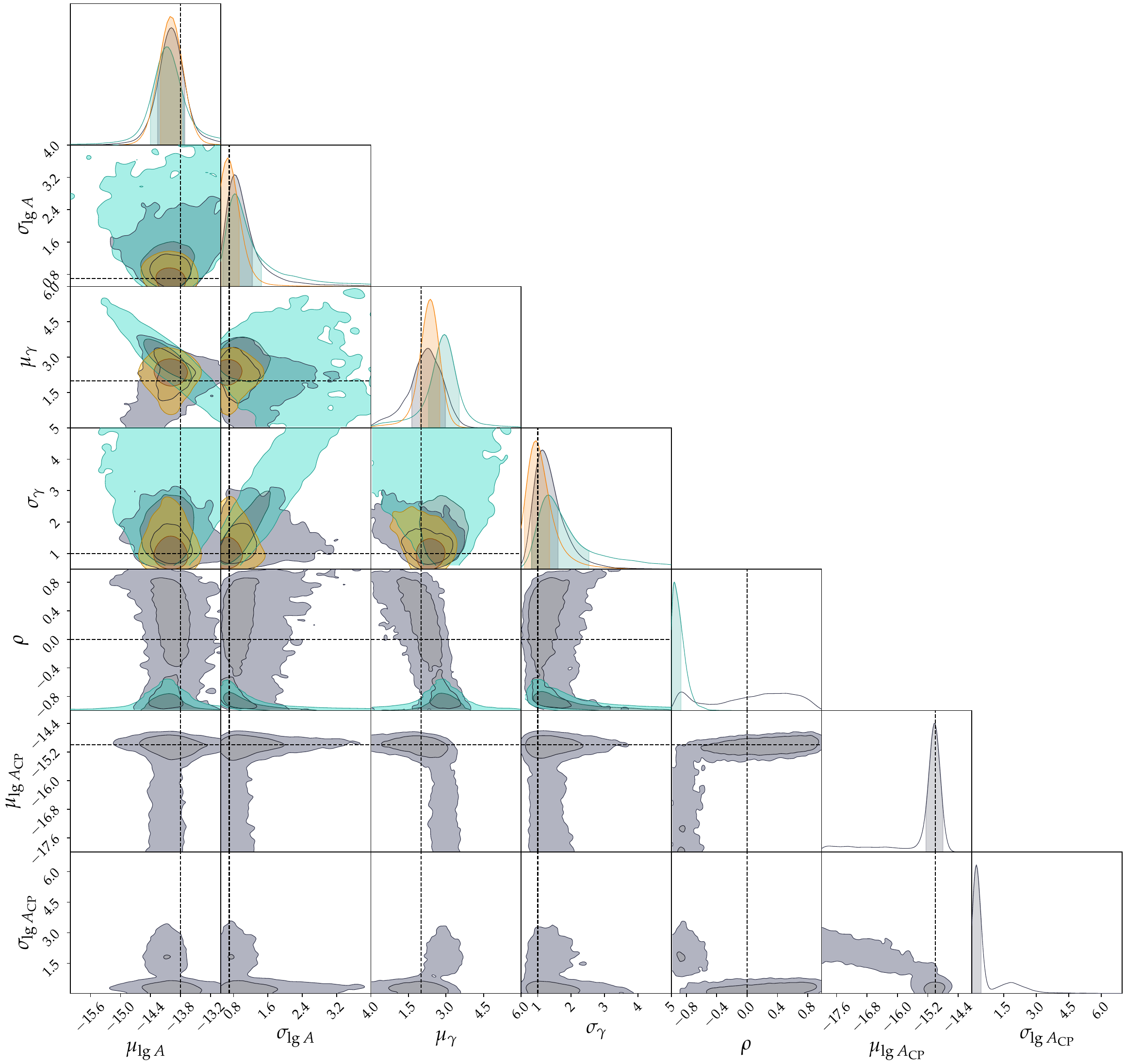}
        %\caption{1}
        \label{fig:quasicommon:sig_lgA2}
    %\end{subfigure}
    %\vspace{-2.2\baselineskip} % to shorten space
    \caption{Posteriors on the inferred hyperparameters in the second simulation. Dashed vertical lines correspond to simulated values in Table \ref{tab:sim}. Colours and contours are explained in Figure \ref{fig:sim1}.
    }
    \label{fig:sim2}
\end{figure*}

First, we perform (hyper-)parameter estimation assuming no covariance between spin noise amplitudes and spectral indices, as simulated ($\rho=0$). 
The results for the first simulation are shown in Figure \ref{fig:sim1}, and for the second simulation in Figure \ref{fig:sim2}, both as orange contours.
The inferred hyperparameters are fully consistent with the simulated values, except for the standard deviation of the spin noise amplitudes $\sigma_{\lg A}$ for the first simulation. 
The simulated value of $1.5$ lies at the edge of the $1\sigma$ credible level in two-dimensional posteriors including $\sigma_{\lg A}$, and just outside of the $1\sigma$ level in the one-dimensional posterior for  $\sigma_{\lg A}$. 

Second, we infer hyperparameters assuming there may be a covariance between spin noise amplitudes and spectral indices ($\rho$ is a free parameter). 
The results are shown as green contours in Figure \ref{fig:sim1} and Figure \ref{fig:sim2}.
Although we simulated $\rho=0$, for both simulations we find $\rho$ to be inconsistent with $0$. 
Instead, the maximum-\textit{aposteriori} $\rho$ is consistent with $-1$, while the value of $0$ is beyond $3\sigma$ credibility. 
Compared to the previous case where we fix $\rho=0$, a minor tension at $1$-$2\sigma$ level emerges for $\mu_\gamma$ in the second simulation, but another minor tension for $\sigma_{\lg A}$ in the first simulation disappears. 
We conclude that $\rho$ is most sensitive to posterior biases. 

To resolve posterior biases, we extend our model to $A_\text{CP}$, imposing that it is also drawn from the truncated Gaussian distribution. 
Thus, we extend both spin noise and CP priors to become a part of our model: $\pi(\lg A, \gamma| \mu_{\lg A}, \sigma_{\lg A}, \mu_{\gamma}, \sigma_{\gamma}, \rho) \pi(\lg A_\text{CP} | \mu_{\lg A_\text{CP}}, \sigma_{\lg A_\text{CP}})$. 
We impose a Log-Uniform prior on $\sigma$ parameters. 
The results are shown as black contours in Figure \ref{fig:sim1} and Figure \ref{fig:sim2}.
In this case, we find the inferred hyperparameters to be fully consistent with the simulated values. 
The values of $\mu_{\lg A_\text{CP}}$ and $\sigma_{\lg A_\text{CP}}$ are also in accordance with our expectations. 
The value $\mu_{\lg A_\text{CP}}$ is consistent with the simulated $\lg A_\text{CP}=-15.0$. 
The value of $\sigma_{\lg A_\text{CP}}$ peaks at small values, it excludes $0$ in the first simulation, despite CP representing an infinitely small range of amplitudes. 
This is a known effect in importance sampling due to a finite number of the recycled posterior samples \citep{GoncharovThrane2022}.

\section{Numerical marginalisation over hyperparameters}\label{sec:marginalisation}

In this Section, we derive Equation~\ref{eq:h_marg} which describes a hierarchical posterior marginalised over hyperparameters $\bm{\Lambda}$. 
After taking an integral over $\bm{\Lambda}$, Equation~\ref{eq:h_likelihood} becomes
\begin{equation}\label{eq:step_1}
    \mathcal{P}(\bm{\theta}|\bm{\delta t}) =  \mathcal{Z}^{-1} \mathcal{L}(\bm{\delta t}|\bm{\theta}) \int  \pi(\bm{\theta}|\bm{\Lambda}) \pi(\bm{\Lambda}) d\bm{\Lambda}.
\end{equation}
Following the formalism of importance sampling which was used to obtain Equation~\ref{eq:prior_reweighting}, we refer to $\pi(\bm{\theta}|\bm{\Lambda})$ as the target distribution. 
Similarly, we introduce the proposal distribution, the standard PTA noise prior that does not depend on hyperparameters, $\pi(\bm{\theta}|\varnothing)$. 
Next, we multiply Equation~\ref{eq:step_1} by unity and expand the unity on the right-hand side as a proposal distribution divided by itself.
Rearranging the multipliers,
\begin{equation}\label{eq:step_2}
    \mathcal{P}(\bm{\theta}|\bm{\delta t}) = \frac{\mathcal{L}(\bm{\delta t}|\bm{\theta}) \pi(\bm{\theta}|\varnothing)}{\mathcal{Z}} \int  \frac{\pi(\bm{\theta}|\bm{\Lambda})}{\pi(\bm{\theta}|\varnothing)} \pi(\bm{\Lambda}) d\bm{\Lambda}.
\end{equation}
Next, we use the expression for the expectation value of a probability density $f(x)$ given the known probability density $p(x)$\footnote{Equation~\ref{eq:integralsum} is also used to derive Equation~\ref{eq:prior_reweighting}. There, $p(x)$ is taken to be a posterior distribution instead.}:
\begin{equation}\label{eq:integralsum}
    \langle f(x) \rangle_{p(x)} = \int f(x) p(x) dx \approx \frac{1}{n_\text{s}} \sum_{i}^{n_\text{s}}f(x_i),
\end{equation}
where $x_i$ are $n_\text{s}$ samples from $p(x)$. 
Taking $\pi(\bm{\Lambda})$ as $p(x)$, we arrive at Equation~\ref{eq:h_marg}.

%%%%%%%%%%%%%%%%%%%%%%%%%%%%%%%%%%%%%%%%%%%%%%%%%%

% Don't change these lines
\bsp	% typesetting comment
\label{lastpage}
\end{document}